\documentclass[aps,prd,showpacs,preprintnumbers]{revtex4}
\usepackage{amsmath,epsfig}
\usepackage{subfigure}
\usepackage{amssymb,amsfonts}
\usepackage{latexsym}
\usepackage{epsfig}
\usepackage{amssymb}
\usepackage{graphicx}
 \usepackage{bm}

\usepackage{color}
\usepackage{hyperref}
\usepackage{textcomp}
\usepackage{wasysym}

\newcommand{\bea}{\begin{eqnarray}}
\newcommand{\eea}{\end{eqnarray}}

\def \be {\begin{equation}}
\def \ee {\end{equation}}
\def \ba {\begin{array}}
\def \ea {\end{array}}
\def \bea{\begin{eqnarray}}
\def \eea{\end{eqnarray}}

\def \a {\alpha}
\def \b {\beta}
\def \g {\gamma}

\def \m {\mu}

\def \k {\kappa}

\def \L {\Lambda}
\def \s {\sigma}

\def \r {\rho}

\def \O {\Omega}

\def \t {\tau}

\def \mL {\mathcal L}

\def \mU {\mathcal U}

\def \p {\partial}
\def \f {\frac}

\def \and {{\textrm{and}}}

\begin{document}

\title{The kurtosis of net baryon number fluctuations from a realistic Polyakov--Nambu--Jona-Lasinio model along the experimental freeze-out line}
\author{Zhibin Li $^{a,b}$}
\thanks{lizb@mail.ihep.ac.cn, co-first author}
\author{Kun Xu $^{a,b}$}
\thanks{xukun@mail.ihep.ac.cn, co-first author}
\author{Xinyang Wang$^{a}$}
\thanks{wangxinyang@mail.ihep.ac.cn}
\author{Mei Huang$^{b,a}$}
\thanks{huangm@ucas.ac.cn, correspondence author}
\affiliation{$^{a}$ Institute of High Energy Physics, Chinese Academy of Sciences, Beijing 100049, China}
\affiliation{$^{b}$ School of Nuclear Science and Technology, University of Chinese Academy of Sciences, Beijing 100049, China}

\begin{abstract}
Firstly we qualitatively analyze the formation of the dip and peak structures of the kurtosis $\kappa \sigma^2$ of net baryon number fluctuation along imagined
freeze-out lines and discuss the signature of the existence of the QCD critical end point (CEP) in the Nambu--Jona-Lasinio (NJL) model, Polyakov-NJL (PNJL) model as well as $\mu$-dependent PNJL($\mu$ PNJL) model with different parameter sets, and then we apply a realistic PNJL model with parameters fixed by  lattice data at zero chemical potential, and quantitatively investigate its $\kappa \sigma^2$ along the real freeze-out line extracted from experiments. The important contribution from gluodynamics to the baryon number fluctuations is discussed. The peak structure of $\kappa \sigma^2$ along the freeze-out line is solely determined by the existence of the CEP mountain and can be used as a clean signature for the existence of CEP. The formation of the dip structure is sensitive to the relation between the freeze-out line and the phase boundary, and the freeze-out line starts from the back-ridge of the phase boundary is required. To our surprise, the kurtosis $\kappa \sigma^2$ produced from the realistic PNJL model along the experimental freeze-out line agrees with BES-I data well, which indicates that equilibrium result can explain the experimental data. It is worth to point out that the extracted freeze-out temperatures from beam energy scan measurement are indeed higher than the critical temperatures at small chemical potentials, which supports our qualitative analysis.
\end{abstract}
\pacs{12.38.Mh,25.75.Nq,11.10.Wx,11.25.Tq }
\maketitle

\section{Introduction}

The phase transition and phase structure of Quantum Chromodynamics (QCD) under extreme conditions is the main topic of relativistic heavy ion collisions, and
it is also highly related to the evolution of the early universe and the equation of state inside the compact stars. Lattice QCD calculation shows that at small baryon
density, the QCD phase transitions including the chiral phase transition as well as deconfinement phase transition are of crossover at finite temperature
\cite{Fodor:2001au,Ding:2015ona,Schmidt:2017bjt}. From symmetry analysis and effective chiral models, it is generally believed  that at high baryon density
the chiral phase transition is of first order and there exists a QCD critical end point (CEP) for chiral phase transition in the temperature and baryon chemical potential
plane \cite{Pisarski:1983ms,Hatta:2002sj,Schwarz:1999dj,Zhuang:2000ub,Chen:2014ufa,Chen:2015dra,Fan:2016ovc,Fan:2017kym,Fu:2010ay,Bowman:2008kc,Mao:2009aq,
Schaefer:2011ex,Schaefer:2012gy,Qin:2010nq,Luecker:2013oda,Fu:2016tey,Stephanov:2008qz,Stephanov:2011pb,Asakawa:2009aj,Athanasiou:2010kw}.
The QCD CEP have been widely analyzed in different models, e.g., Nambu--Jona-Lasinio (NJL) model, the Polyakov-loop improved NJL
(PNJL) model, linear sigma model, quark-meson (QM) model, the Polyakov-loop improved QM model, the Dyson-Schwinger equations (DSE), and the holographic QCD model
\cite{Schwarz:1999dj,Zhuang:2000ub,Chen:2014ufa,Chen:2015dra,Fan:2016ovc,Fan:2017kym,Fu:2010ay,Bowman:2008kc,Mao:2009aq,
Schaefer:2011ex,Schaefer:2012gy,Qin:2010nq,Luecker:2013oda,Fu:2016tey,Critelli:2017oub}. However, different models even the same model with
different parameter sets give various location of CEP\cite{Vovchenko:2017gkg}. Therefore, to search for the existence of the CEP and further to locate the CEP is one of the most central
goals at Relativistic Heavy Ion Collisions (RHIC) as well as for the future accelerator facilities at Facility for Antiproton and Ion Research (FAIR) in
Darmstadt and Nuclotron-based Ion Collider Facility (NICA) in Dubna.

The cumulants of conserved quantities up to fourth order of net-proton, net-charge and net-kaon multiplicity distributions have been measured in the first phase of
beam energy scan program (BES-I) at RHIC for Au+Au collisions at $\sqrt{s_{NN}}=7.7, 11.5, 14.5, 19.6, 27, 39, 62.4$ and $200 {\rm GeV}$, and the
results are summarized in \cite{Adamczyk:2013dal,Aggarwal:2010wy,Luo:2017faz}. A non-monotonic energy dependent behavior for the kurtosis of the net proton
number distributions $\kappa \sigma^2$ has been observed in the most central Au+Au collisions: $\kappa \sigma^2$ firstly decreases from around 1 at the colliding
energy $\sqrt{s_{NN}}=200 {\rm GeV}$ to 0.1 at $\sqrt{s_{NN}}=20{\rm GeV}$ and then rises quickly up to around 3.5 at $\sqrt{s_{NN}}=7{\rm GeV}$.
From the BES-I observed oscillation behavior of kurtosis, we need to know what information we can extract about the CEP and where is the CEP located. Before the second phase of beam energy scan (BES-II) at RHIC performed in 2019-2020, it is very urgent for both experimentalists and theorists to extract a clean signature to identify the existence of the QCD CEP and further to locate the CEP. On the one hand, we should try to extract useful information about QCD phase transitions from the measurement along the freeze out line, and to find the evidence of the QCD CEP. On the other hand, we should explore carefully how QCD phase transitions will shed light on properties of the cumulants of conserved quantities at freeze-out.

By using the Ising model, it was pointed out in \cite{Stephanov:2011pb} that the quartic cumulant (or kurtosis) of the order parameter is universally negative when the
critical point is approaching to the crossover side of the phase separation line. From the results in \cite{Stephanov:2011pb}, the oscillation behavior of the kurtosis along the freeze out line has been regarded as a typical signature of the existence of the CEP. Many interests have focused on the sign changing of various cumulants
around the CEP, and the sign changing for higher order susceptibilities has been recently discussed in \cite{Fan:2017kym}. It is noticed that
the sign changing for the 6th and 8th order susceptibilities starts at the baryon chemical potential quite far away from the CEP,  it may indicate that the sign changing
of cumulans of conserved quantities is not directly related to the CEP. In the holographic QCD model \cite{Li:2017ple}, we explained that the sign changing
of the baryon number susceptibilities along the freeze-out line is not necessarily related to the CEP, but the peaked baryon number susceptibilities along the freeze-out
line is solely determined by the CEP thus can be used as an evident signature for the existence of the CEP, and the peak position is close to the location of the CEP in the QCD phase diagram. Furthermore, it is found that at zero chemical potential, the magnitude of $\kappa \sigma^2$ around the phase transition line in the gluodynamics dominant holographic QCD model  is around $1$, which is in agreement with lattice result, and is much larger than that in the Ising model (close to zero), and also much larger than that in the NJL model (around 0.1). Therefore, it is natural to speculate that the gluodynamics contribution is dominant to the baryon number
susceptibilities, which is quite surprising!

In order to check the gluodynamics contribution to the baryon number susceptibilities, in this work, we will try to explore the structure of the kurtosis of the net
proton number distribution $\kappa \sigma^2$ along the freeze out line in the Nambu--Jona-Lasinio (NJL) model, the Polyakov-loop improved NJL (PNJL) model
as well as $\mu$-dependent Polyakov-loop potential improved NJL ($\mu$PNJL) model. Also in order to investigate the relation between the CEP and the structure of the $\kappa \sigma^2$ along the freeze out line, we will need to shift the location of the CEP by introducing the interaction in the vector channel. This paper is organized as following: After Introduction, in Sec.  \ref{sec-qualitative-model}, we give a brief introduction to the two-flavor NJL model,  the Polyakov-loop improved NJL (PNJL) model as well as $\mu$ Polyakov-loop improved NJL ($\mu$PNJL) model, and qualitatively analyze the formation of the dip and peak structures of the kurtosis $\kappa \sigma^2$ of net baryon number fluctuation along imagined freeze-out lines and discuss the signature of the existence of the QCD critical end point (CEP).  Then in Sec. \ref{sec:realistic-model} we apply a realistic PNJL model with parameters fixed by  lattice data at zero chemical potential, and quantitatively investigate its $\kappa \sigma^2$ along the real freeze-out line extracted from experiments. Finally, the discussion and conclusion part is given in Sec.~\ref{sec:Conclusion}.

\section{Qualitative analysis of baryon number fluctuations in the NJL, PNJL and $\mu$PNJL Models}
\label{sec-qualitative-model}

In order to qualitatively analyze the formation of the dip and peak structure of the kurtosis of the net baryon number distribution $\kappa \sigma^2$ along the freeze-out line, as well as to investigate the gluodynamics contribution to $\kappa \sigma^2$, in this section we will compare $\kappa \sigma^2$
in the framework of NJL model, the Polyakov-loop improved NJL (PNJL) model as well as NJL with $\mu$-dependent Polyakov-loop potential ($\mu$PNJL) model. For each model, except the coupling constant in the vector channel, the parameters from quark part are fitted by vacuum properties as well as pion mass and decay constant, and the parameters from the Polykov-loop potential part are fitted from the Lattice results of the equation of state at $\mu=0$. We will intendedly shift the location of the CEP in the models by changing the coupling constant in the vector channel, to check the peak structure of $\kappa \sigma^2$  along the imagined freeze-out lines and its relation with the location of CEP.

\subsection{The NJL, PNJL and $\mu$PNJL models with vector interaction}

In order to shift the location of the CEP, we introduce the two-flavor NJL model with the vector interaction, and the Lagrangian is given by
\cite{Nambu:1961tp,Nambu:1961fr,Klevansky:1992qe}
\be
\mL_{\text{NJL}}=\bar{\psi}(i\g_\m \partial^\m-m)\psi+G_S [(\bar{\psi} \psi)^2+(\bar \psi i\g_5\vec{\t} \psi)^2]
-G_V [(\bar{\psi}\g_\m \psi)^2+(\bar{\psi}\g_\m\g_5 \psi)^2].
\ee
Where $\psi=(u,d)^T$ is the doublet of the two light quark flavors $u$ and $d$ with the current mass $m=m_u=m_d$, and $\vec{\tau}=(\tau^1,\tau^2,\tau^3)$
the isospin Pauli matrix. $G_S$ and $G_V$ are the coupling constants in the (psudo)scalar channel and the vector channel, respectively.
By introducing the auxiliary fields for scalars and vectors, and in the vacuum we take the mean-field approximation with
\be
\s_i=\langle \bar{\psi}_i \psi_i \rangle, \hspace{0.3cm} \r_i=\langle \psi^\dagger_i \psi_i \rangle,
\ee
the quark condensate and the net quark number density of flavor $i$ respectively.
Then the thermodynamical potential of this NJL model takes the following form:
\be
\O_{\text{NJL}}=-2N_c\sum_{i=u,d}\int_0^\L \f{d^3p}{(2\pi)^3}[E_i+T \ln(1+e^{-\b(E_i-\tilde{\m}_i)})+T\ln(1+e^{-\b(E_i+\tilde{\m}_i)})]
+G_S( \s_u+ \s_d)^2-G_V(\r_u+\r_d)^2,
\ee
with $N_c=3$ the number of colors. The quark quasiparticle energies $E_i$ and constituent quark masses $M_i$ for flavors $i=(u,d)$ are given by
\be
E_i=\sqrt{p^2+M_i^2},\hspace{0.3cm} M_i=m_i-2G_S( \s_u+\s_d),
\ee
and the effective chemical potentials are shifted by
\be
\tilde{\m}_i=\m_i-4 G_V \r_i.
\ee
In order to solve the minimum of the thermal potential $\O_{\text{NJL}}$, we have the following gap equations
\be
\f{\p \O_{NJL}}{\p \s_u}=\f{\p \O_{NJL}}{\p \s_d}=0,
\ee
and
\be
\r_i=-\f{\p \O_{NJL}}{\p \m_i}.
\ee
For numerical calculations, we fix $m_u=m_d=5.5{\rm MeV}, N_c=3, N_f=2$ in all the models. In the regular NJL model, we choose the parameters as in Ref.\cite{Dutra:2013lya} by fitting the pion mass and decay constant. To shift the location of the CEP in the NJL model we choose different coupling constant in the vector channel: $G_V=-0.5 G_S$, $G_V=0$ and $G_V=0.67 G_S$ \cite{Fan:2017kym}, which is indicated as NJL-1, NJL-2, NJL-3 respectively, and three sets of parameters are shown in Table. \ref{table-NJL-parameters}. Correspondingly, the locations of CEP in the NJL-1 and NJL-2 are $(\mu_B^E=796.7 {\rm MeV},T^E=76.8 {\rm MeV})$ and $(\mu_B^E=1005.2 {\rm MeV},T^E=34.6 {\rm MeV})$, and there is no CEP in the NJL-3.

\begin{table}[!ht]\centering
\begin{tabular}{c|c|c|c|c|c}\hline\centering
	 &  $\Lambda ({\rm MeV})$ & $G_S ({\rm GeV}^{-2})$  & $G_V/G_S$  & $ T^E ({\rm MeV})$  & $\mu_B^E({\rm MeV}) $ \\\hline
  NJL-1   &   651  &    5.04    &    -0.5     & 76.8  &796.7   \\ \hline
   NJL-2  &   651  &    5.04    &      0       & 34.6 & 1005.2  \\ \hline
   NJL-3  &   651  &    5.04    &      0.67   &  No &  No \\ \hline
\end{tabular}
\caption{Three sets of parameters used in the NJL model, and the corresponding critical temperatures and chemical potentials at the critical end point.}
\label{table-NJL-parameters}
\end{table}


In the 2-flavor PNJL and $\m$PNJL models, we need to add the Polyakov-loop effective potential $\mU(\Phi, \bar{\Phi}, T)$ with the following ansatz \cite{Fukushima:2008wg,Weiss:1980rj,Ratti:2006wg,Shao:2017yzv}
\be
\f{\mU(\Phi, \bar{\Phi}, T)}{T^4}=-\f{a(T)}{2} \bar{\Phi}\Phi+b(T)\ln[1-6\bar{\Phi}\Phi+4(\bar{\Phi}^3+\Phi^3)-3(\bar{\Phi}\Phi)^2],
\ee
where $a(T)=a_0+a_1(\f{T_0}T)+a_2(\f{T_0}T)^2$ and $b(T)=b_3(\f{T_0}T)^3$. The critical temperature of the confinement-deconfinement phase transition
$T_0$ in pure gluon system is $270 {\rm MeV}$, and will be rescaled to about $220 {\rm MeV}$ because of the presence of fermion fields.
The thermodynamical potential of this PNJL model is given by
\bea
 \O_{PNJL}&=&\mU(\Phi, \bar{\Phi}, T)-2N_c\sum_{i=u,d}\int_0^\L \f{d^3p}{(2\pi)^3}[E_i]+G_S( \s_u+ \s_d)^2-G_V(\r_u+\r_d)^2 \nonumber \\
 & & -2T\sum_{i=u,d}\int \f{d^3p}{(2\pi)^3}[\ln(1+3\Phi e^{-\b(E_i-\tilde{\m}_i)}
 +3\bar\Phi e^{-2\b(E_i-\tilde{\m}_i)}+e^{-3\b(E_i-\tilde{\m}_i)})] \nonumber \\
& &-2T\sum_{i=u,d}\int \f{d^3p}{(2\pi)^3}[\ln(1+3\bar{\Phi} e^{-\b(E_i+\tilde{\m}_i)} +3\Phi e^{-2\b(E_i+\tilde{\m}_i)} +e^{-3\b(E_i+\tilde{\m}_i)})]
\eea
with the same definitions of $E_i$, $M_i$ and $\tilde\m_i$ as in the NJL model.
The gap equations are determined by
\be
\f{\p \O_{PNJL}}{\p \s_u}=\f{\p \O_{PNJL}}{\p \s_d}=\f{\p \O_{PNJL}}{\p \Phi}=\f{\p \O_{PNJL}}{\p \bar\Phi}=0,
\ee
and
\be
\r_i=-\f{\p \O_{PNJL}}{\p \m_i}.
\ee
In the PNJL model,  we fix $T_0=270 {\rm MeV}$,  and choose the parameter sets by fitting the experimental values of pion decay constant $f_\pi=92.3{\rm MeV}$ and the pion mass $m_\pi=139.3{\rm MeV}$ when $G_V=0$. And we choose different $G_V$ for PNJL-1, PNJL-2 and PNJL-3 as shown in Table.\ref{Table-parameters-PNJL}. For the PNJL-1, at zero chemical potential $\mu=0$, the critical temperature for the
chiral phase transition is $T_0^{\chi}=222.9{\rm MeV}$, and for deconfinement phase transition is $T_0^D=214.3 {\rm MeV}$, and the CEP is located at $(\mu_B^E=919.1{\rm MeV},T^E=123.5{\rm MeV})$. For the PNJL-2, at $\mu=0$, the critical temperature for the
chiral phase transition is $T_0^{\chi}=223.6{\rm MeV}$, and for deconfinement phase transition is $T_0^D=215.0 {\rm MeV}$, and the CEP is located at
$(\mu_B^E=979.5{\rm MeV}, T^E=104.2{\rm MeV})$. For the PNJL-3, at $\mu=0$, the critical temperature for the
chiral phase transition is $T_0^{\chi}=222.4{\rm MeV}$, and for deconfinement phase transition is $T_0^D=214.8 {\rm MeV}$,
and there is no CEP in the PNJL-3.

\begin{table}[!ht]\centering
\begin{tabular}{c|c|c|c|c|c|c|c|c|c}\hline\centering
	& $\Lambda ({\rm MeV})$ & $G_S $  & $G_V/G_S$  & $a_0$   & $a_1$ & $a_2$ & $b_3$ & $ T^E ({\rm MeV})$  & $\mu_B^E({\rm MeV}) $ \\\hline
  PNJL-1   &	 651    &   5.04   &    -0.15 &    3.51          &      -2.47      & 15.22    &  -1.75 & 123.5 & 919.1 \\ \hline
   PNJL-2  & 651        &   5.04   &    0 &    3.51          &      -2.47       &   15.22  &  -1.75 & 104.2 & 979.5 \\ \hline
   PNJL-3  & 651        &   5.04    &    0.67&    3.51         &      -2.47     &   15.22  &  -1.75& No & No \\ \hline
\end{tabular}
\caption{Three sets of parameters used in the PNJL model, and the corresponding critical temperatures and chemical potentials at the critical end point.}
\label{Table-parameters-PNJL}
\end{table}

By considering the back-reaction of quark to the glue sector, we take the so called $\m$PNJL model with the $\m$-dependent
$T_0$\cite{Schaefer:2007pw,Xin:2014dia,Shao:2016fsh}
\be
T_0(N_f,\m_i)=T_\t e^{-\f 1 {\a_0 f(N_f,\m_i)}}
\ee
where $f(N_f,\m_i)=\f{11 N_c-2N_f}{6\pi}-\k\f{16N_f}\pi\f{\m^2}{T_\t^2}$ and $\m=\f 1{N_f}\sum_{i} \m_i$.
The thermodynamical potential of $\m$PNJL model $\O_{\m\text{PNJL}}$ has the same form of PNJL model, and the gap equations take the following form:
\be
\f{\O_{\m\text{PNJL}}}{\p \s_u}=\f{\p \O_{\m\text{PNJL}}}{\p \s_d}=\f{\p\O_{\m\text{PNJL}}}{\p \Phi}=\f{\p\O_{\m\text{PNJL}}}{\p \bar\Phi}=0,
\ee
and
\be
\r_i=-\f{\p \O_{\m\text{PNJL}}}{\p \m_i}.
\ee
In the $\m$PNJL model, we also choose three different parameter sets as shown in Table \ref{Table-parameters-muPNJL}\cite{Shao:2016fsh}.
For the $\mu$PNJL-1, at $\mu=0$, the critical temperature for the
chiral phase transition is $T_0^{\chi}=201.4{\rm MeV}$, and for deconfinement phase transition is $T_0^D=193.7 {\rm MeV}$, and the CEP is located at
$(\mu_B^E=909.5{\rm MeV},T^E=121.6{\rm MeV})$.
For the $\mu$PNJL-2, at $\mu=0$, the critical temperature for the
chiral phase transition is $T_0^{\chi}=202.4{\rm MeV}$, and for deconfinement phase transition is $T_0^D=192.8 {\rm MeV}$, and the CEP is located at
$(\mu_B^E=975.7{\rm MeV},T^E=93.4{\rm MeV})$.
For $\m$PNJL-3, at $\mu=0$, the critical temperature for the
chiral phase transition is $T_0^{\chi}=202.5{\rm MeV}$, and for deconfinement phase transition is $T_0^D=192.7 {\rm MeV}$, and there is no CEP in the $\m$PNJL-3.

\begin{table}[!ht]\centering
\begin{tabular}{c|c|c|c|c|c|c|c|c|c|c|c|c}\hline\centering
	& $\Lambda ({\rm MeV})$ & $G_S $ &  $G_V/G_S$ & $T_{\tau}$  & $\alpha_0$& $\kappa$  & $a_0$   & $a_1$ & $a_2$ & $b_3$ & $ T^E ({\rm MeV})$  & $\mu_B^E({\rm MeV}) $ \\\hline
  $\mu$PNJL-1   &	 651 & 5.04  &   -0.15  &    1770 &   0.304 & 0.1&  3.51     &      -2.47    & 15.2    &  -1.75 & 121.6 & 909.5 \\ \hline
  $\mu$PNJL-2  & 651    &  5.04  &    0 &    1770 &    0.304 & 0.1    &3.51 &      -2.47    &   15.2 &  -1.75 & 93.4 & 975.7 \\ \hline
  $\mu$PNJL-3  & 651    &  5.04  &   0.67 &    1770 &    0.304 & 0.1    &3.51 &      -2.47    &   15.2 &  -1.75 & NO & NO \\ \hline
\end{tabular}
\caption{Three parameters sets used in the $\mu$PNJL model, and the corresponding critical temperatures and chemical potentials at the critical end point.}
\label{Table-parameters-muPNJL}
\end{table}

\subsection{The gluodynamics contribution to $\kappa \sigma^2$ }

BES-I measures the net-proton number susceptibilities, which can be approximately regarded as net-baryon number fluctuations and are defined as the derivative of the dimensionless pressure with respected to the reduced chemical potential \cite{Luo:2017faz}
\be
\chi^B_n=\f{\p ^n[P/T^4]}{\p [\m_B /T]^n},
\ee
with the pressure $P=-\Omega$ which is just the minus thermodynamical potential.
The cumulants of baryon number distributions are given by
\be
C_n^B=VT^3\chi^B_n.
\ee
By introducing the variance $ \s^2=C_2^B$,  kurtosis $\k=\f{C_4^B}{(\s^2)^2}$, one can have the following relation between observable quantities
and theoretical calculations:
\begin{eqnarray}
 \k \s^2=\f{C_4^B}{C_2^B}=\f{\chi_4^B}{\chi_2^B},
\label{Eq:ratios}
\end{eqnarray}
which relates $\k \s^2$ with the ratio of fourth and second order cumulants of net-baryon number fluctuations.

Because the measurement of $\k \s^2$ from BES-I is along the freeze-out line, here we define three different imagined freeze-out lines for the NJL, PNJL and $\mu$PNJL models as:
\bea\label{eq:freezeout}
&& (\rm{p}, \mu \rm{p})f\rm{I}1:\hspace{2mm}T_1(\mu_B)=1.03 T^I_p(\mu_B)-10^{-5}\mu_B^2-10^{-10}\mu_B^{4} \nonumber\\
&& (\rm{p}, \mu \rm{p})f\rm{I}2:\hspace{2mm}T_2(\mu_B)= T^I_p(\mu_B)-3\times 10^{-5}\mu_B^2-3\times 10^{-10}\mu_B^{4} \nonumber\\
&& (\rm{p}, \mu \rm{p})f\rm{I}3:\hspace{2mm}T_3(\mu_B)=0.96 T^I_p(\mu_B)-8\times 10^{-5}\mu_B^2-6\times 10^{-10}\mu_B^{4}
\eea
Here $\rm{I}=1, 2, 3$ is the label of different model e.g. NJL (PNJL or $\mu$PNJL)-I and $\rm{p}, \mu \rm{p}$ is for the PNJL and $\mu$PNJL model, respectively. And $T^I_p(\mu_B)$ is the function fitting of chiral phase transition line and both $T$ and $\mu_B$ are in $\rm{MeV}$s. These three imagined freeze-out lines just qualitatively indicate how far the freeze-out line is away from the phase boundary, which is an important information to analyze the structure of $\k \s^2$ along the freeze-out line.

\begin{figure}[!thb]
  \centering
  \includegraphics[width=0.95\textwidth]{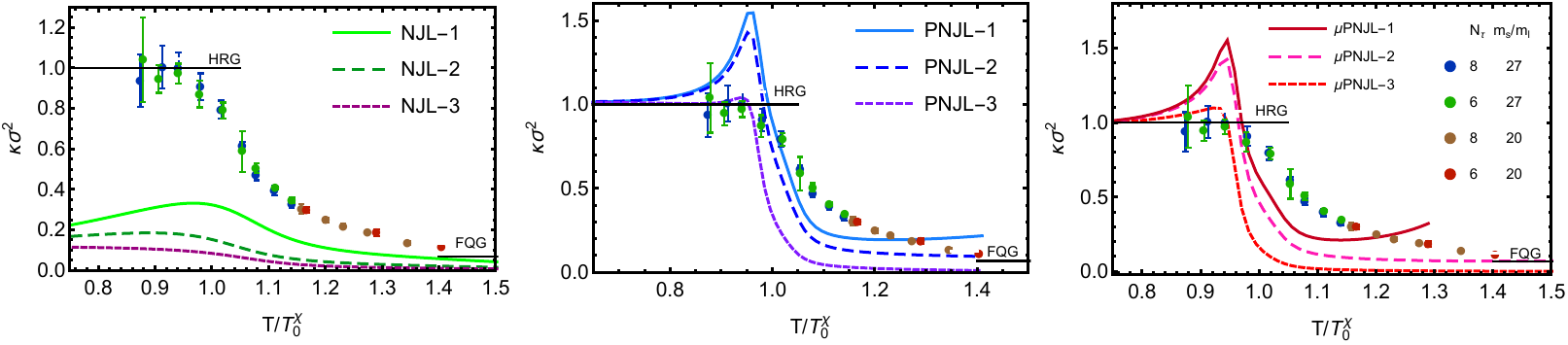}\\
  \caption{The kurtosis of baryon number fluctuation $\kappa \sigma^2$ as a function of the temperature at zero baryon chemical potential in the NJL model, PNJL model and $\mu$PNJL model, and compare with the lattice result in \cite{Bazavov:2017dus}. $\kappa \sigma^2=1$ in the hadron resonance gas (HRG) limit and $\kappa \sigma^2\simeq 0.068$  in the ideal free quark gas (FQG) limit.
}
  \label{fig:withlattice}
\end{figure}

In Fig.\ref{fig:withlattice}, we show $\kappa \sigma^2$ as a function of $T/T^\chi_0$ in three parameter sets of NJL model, and three parameter sets
of PNJL model and three parameter sets of $\mu$PNJL model at zero chemical potential, and compare with lattice results in \cite{Bazavov:2017dus},
with $T_0^\chi$ the critical temperature for chiral phase transition at zero chemical potential $\mu_B=0$. The critical temperature of lattice data at $\m_B=0$
is $154\pm 9{\rm MeV}$ and we choose $154{\rm MeV}$. The value of $\kappa \sigma^2$ of net baryon number fluctuations is unity in the limit of
hadron resonance gas (HRG) and, in the ideal free quark gas (FQG) limit at infinite temperature it takes the value of $\kappa \sigma^2\simeq 0.068$, which are also
shown in Fig.\ref{fig:withlattice}. It is worthy of mentioning that from the lattice result shown in Ref.\cite{Bazavov:2017dus}, that at the critical temperature
$T_0^\chi$, the magnitude of $\kappa \sigma^2$ is around 0.8, which is smaller than its value of 1 at the temperature of $T_0^{dec}\simeq 140 {\rm MeV}$, at which
quark matter transfers to hadron gas. Here $T_0^{dec}$ describes a physical "confinement-deconfinemen" transition, which is different from the $T_0^D$ describing the confinement-deconfinement phase transition by using the order parameter of Polyakov loop.

In the NJL model, the magnitude of $\kappa \sigma^2$ at $T_0^\chi$ is around $0.4,0.2,0.1$ in the NJL-1, NJL-2 and NJL-3, which is much smaller
than the lattice result. In the whole temperature region $T>T_0$, the magnitude of $\kappa \sigma^2$ is quite small comparing with the lattice results.

When the gluodynamics contribution is taken into account in the PNJL model and $\m$PNJL model, the critical temperatures for the chiral phase transition $T_0^{\chi}$
and deconfinement phase transition $T_0^{D}$ are separated, and the critical temperature for the deconfinement phase transition $T_0^{D}$ is lower than that
of the chiral phase transition $T_0^{\chi}$ at zero chemical potential. It is observed that the magnitude of $\kappa \sigma^2$ shows a peak at the critical temperature for deconfinement phase transition $T_0^D$, and in the hadron gas phase with $T<T_0^D$, the magnitude of $\kappa \sigma^2$ decreases with the decreasing temperature, and matches the HRG limit at low temperature.

From the magnitude of  $\kappa \sigma^2$ in the NJL model, PNJL model and $\mu$PNJL model, we can see that the dominant contribution to $\kappa \sigma^2$
comes from gluodynamics at zero chemical potential. In the future for model construction, $\kappa \sigma^2$ as a function of the temperature can be used to
constrain models.

\subsection{The kurtosis of baryon number fluctuation $\kappa \sigma^2$ in the NJL model}

\begin{figure}[!thb]
\centering
\includegraphics[width=0.9\textwidth]{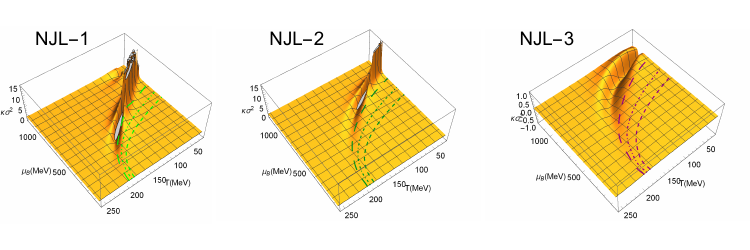}\\
\caption{The 3D plot for the kurtosis of the baryon number fluctuation $\kappa \sigma^2$ as a function of the temperature and baryon chemical potential
  in the NJL model. The long dashed, dashed-dotted and dashed lines in each model NJL-I (with I=1,2,3) are imagined freeze-out lines fI1, fI2 and fI3 defined in Eq.(\ref{eq:freezeout}), respectively.}
\label{fig:Fig2-3DNJL}
\end{figure}

\begin{figure}[!thb]
\centering
\includegraphics[width=0.9\textwidth]{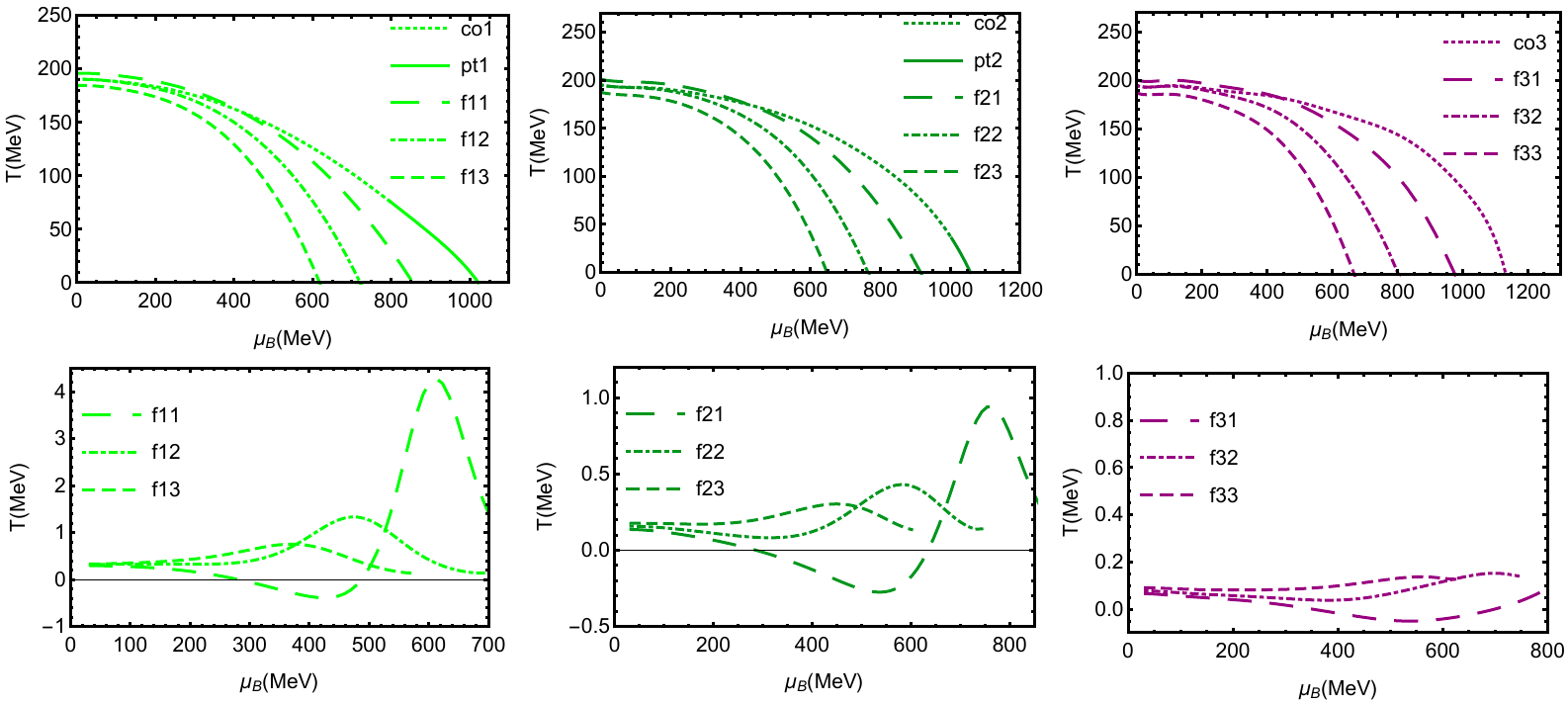}\\
\caption{(Above) The chiral phase transition line in the NJL model with the 1st order phase transition line and the crossover indicated by solid and dotted lines, respectively. Three different freeze-out lines  for each model NJL-I (with I=1,2,3) are defined in Eq.(\ref{eq:freezeout}): 1) fI1 starts from the back ridge of the
chiral phase boundary, goes through the negative region and then cross the phase boundary;
2) fI2 starts from the chiral phase boundary and then cross the foot of the CEP mountain;
3) fI3 is far away from the phase boundary. These three freeze-out lines are indicated by long dashed, dashed-dotted and dashed lines, respectively.
(Below) The 2D plot for $\kappa \sigma^2$ as a function of the baryon chemical potential along three freeze-out lines.}
\label{fig:Fig3-2DNJL}
\end{figure}

In Fig. \ref{fig:Fig2-3DNJL}, we show the 3D plot for the kurtosis of baryon number fluctuation $\kappa \sigma^2$ as a function of the temperature and baryon
chemical potential in the NJL model. In Fig.\ref{fig:Fig3-2DNJL} we show the chiral phase transition line and 2D plot for $\kappa \sigma^2$ as a function of the
baryon chemical potential along different freeze-out lines for NJL-1,NJL-2 and NJL-3, respectively.

By changing the coupling constant in the vector channel, the location of the CEP will shift. The CEP in the NJL-1 and NJL-2 is located at
$(\mu_B^E=796.7{\rm MeV},T^E=76.8{\rm MeV})$, $(\mu_B^E=1005.2{\rm MeV},T^E=34.6{\rm Me}V)$, respectively,  and
it is observed that $ \k \s^2$ develops a high CEP mountain in the NJL-1 and NJL-2. There is no CEP in the NJL-3, therefore, no CEP
mountain of $ \k \s^2$ develops in the NJL-3. From the 3D plot in Fig. \ref{fig:Fig2-3DNJL}, one can observe an obvious chiral phase
boundary in the crossover side, and the magnitude of its ridge decreases with the increase of the baryon chemical
potential. Here we define the ridge as along the phase boundary, the back/front ridge as the higher/lower temperature side
comparing with the phase boundary, respectively. If there is a CEP located on the phase diagram, a CEP mountain rises up around the CEP, and
one can observe a negative region of $\k \s^2$ around the CEP above the chiral phase boundary extended from the crossover side.

Because the measurement of heavy-ion collision is along the freeze-out line, we choose three different freeze-out lines in the NJL model defined in Eq.(\ref{eq:freezeout}):
1) Starting from the back ridge of the chiral phase boundary, goes through the negative region and then cross the phase boundary;
2) Starting from the chiral phase boundary and then cross the foot of the CEP mountain;
3) Far away from the phase boundary. These three freeze-out lines are indicated by long dashed,
dashed-dotted and dashed lines, respectively.

It is observed that when the freeze-out line crosses the CEP mountain or crosses the foot of the CEP mountain, there will be a peak showing up for $ \k \s^2$
along the freeze-out line, and the location of the peak is close to the location of the CEP mountain. In the case of no CEP, it is found that $ \k \s^2$ keeps flat
in almost the whole region and does not show any structure. In the NJL-1 and NJL-2 with the existence of CEP in the phase diagram, it is found that for the first case freeze-out line, because the freeze-out line starts from the back ridge of the chiral phase boundary and goes through the negative region of $\k \s^2$, then crosses the CEP mountain, therefore,
$ \k \s^2$ decreases from around 0.2 and then down to negative value then rises up quickly and shows up a high peak around the critical chemical potential,
which shows a dip and then a peak structure of $ \k \s^2$ along the freeze-out line. For the second case of freeze-out line, because the freeze-out line starts from
the chiral phase boundary, and then crosses the foot of the CEP mountain, we can only see a peak structure of $ \k \s^2$ along the freeze-out line
at high chemical potential. For the third case of freeze-out, if the freeze-out line is far away from the phase boundary, one can only observe a weak peak of $ \k \s^2$ along the freeze-out line.

It is found that the magnitude of $ \k \s^2$ in the NJL model at small baryon chemical potential region is small comparing with experiment measurement which is around 1.

\subsection{The kurtosis of the baryon number fluctuation $\kappa \sigma^2$ in the PNJL model}

\begin{figure}[!thb]
  \centering
  \includegraphics[width=0.9\textwidth]{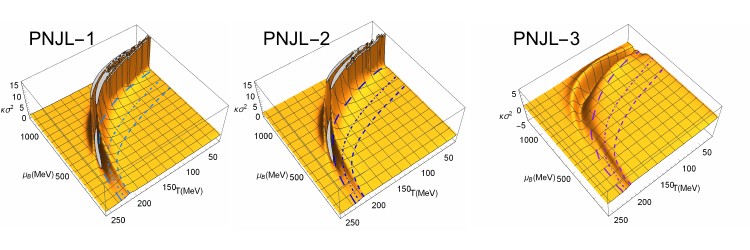}\\
  \caption{The 3D plot for the kurtosis of baryon number fluctuation $\kappa \sigma^2$ as a function of the temperature and baryon chemical potential in the
  PNJL model. The long dashed, dashed-dotted and dashed lines in each model PNJL-I (with I=1,2,3) are imagined freeze-out lines pfI1, pfI2 and pfI3 defined in Eq.(\ref{eq:freezeout}), respectively. }
  \label{fig:Fig4-3DPNJL}
\end{figure}

\begin{figure}[!thb]
  \centering
  \includegraphics[width=0.9\textwidth]{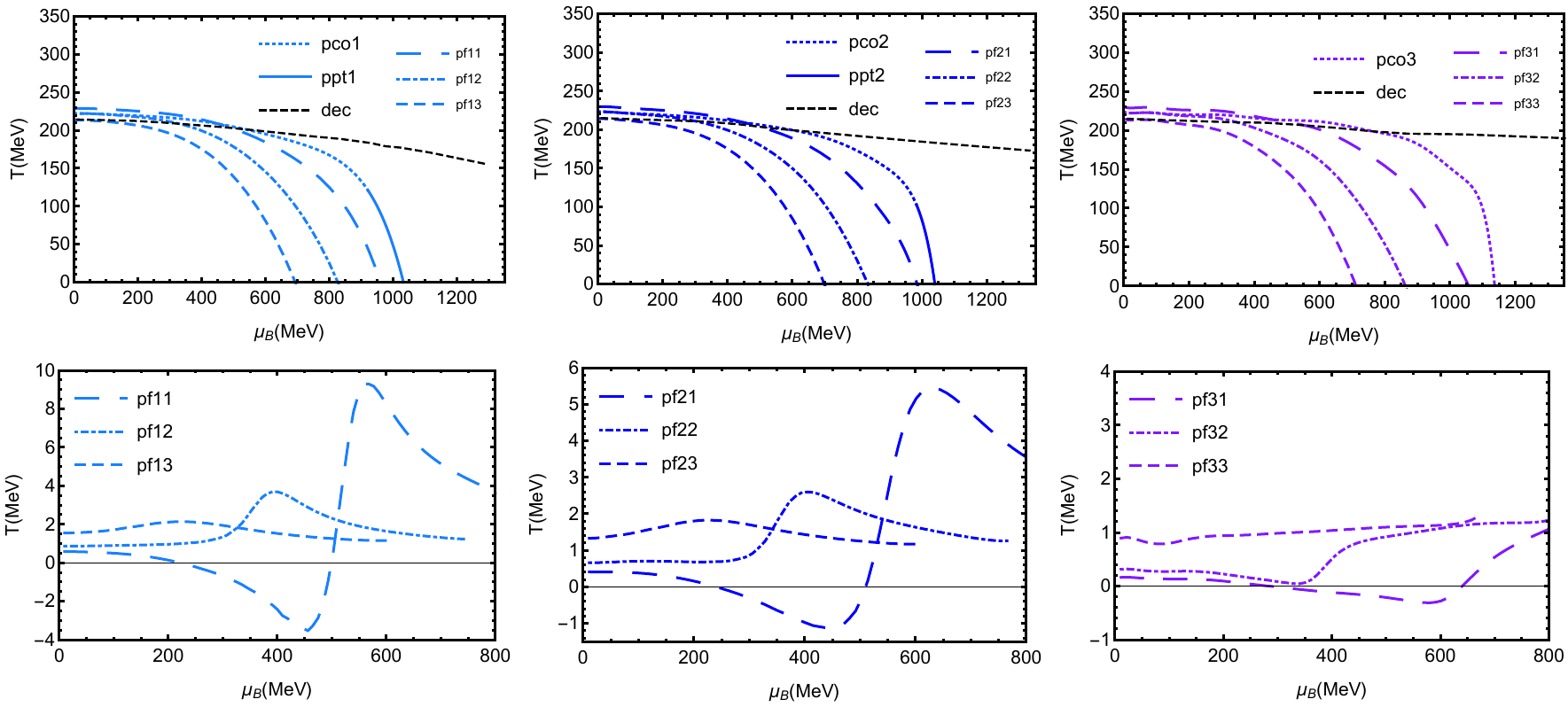}\\
  \caption{(Above) The deconfinement phase transition line indicated by short-dashed line and the chiral phase transition line in the PNJL model with the 1st order phase transition line and the crossover indicated by solid and dotted lines, respectively. Three different freeze-out lines  for each model PNJL-I (with I=1,2,3) are defined in Eq.(\ref{eq:freezeout}):  1) pfI1 starts from the back ridge of the chiral phase boundary, goes through the negative region and then crosses the foot of the CEP mountain;  2) pfI2 starts from the back ridge of the deconfinement phase boundary, and then crosses the foot of the CEP mountain;  3) pfI3 starts  from the deconfinement phase boundary
and keeps far away from the CEP mountain. These three different freeze-out lines are indicated by long dashed, dashed-dotted and
dashed lines, respectively. (Below) The 2D plot for $\kappa \sigma^2$ as a
function of the baryon chemical potential in the PNJL model along three freeze-out lines.}
  \label{fig:Fig5-2DPNJL}
\end{figure}

In Fig. \ref{fig:Fig4-3DPNJL}, we show the 3D plot for the kurtosis of baryon number fluctuation $\kappa \sigma^2$ as a function of the temperature and baryon
chemical potential in the PNJL model. In Fig.\ref{fig:Fig5-2DPNJL} we show the phase transition lines and 2D plot for $\kappa \sigma^2$ as a function of the
baryon chemical potential along different freeze-out lines for PNJL-1,PNJL-2 and PNJL-3, respectively.

Different from the NJL model, in the PNJL model, when the gluodynamics is taken into account, there will be two phase transitions: one for the chiral phase transition
and another for the deconfinement phase transition, and the deconfinement phase transition line lays below the chiral phase transition line at small chemical potential region.
The two separate phase transition lines can be obviously seen from the $(T,\mu)$ phase diagram in Fig.\ref{fig:Fig5-2DPNJL}. In the 3D plot Fig.\ref{fig:Fig4-3DPNJL}
one can observe two separate phase boundaries at small baryon chemical potentials, and a valley forms in between the two phase boundaries. It is noticed that the
magnitude of baryon number fluctuation $\kappa \sigma^2$ is quite small along the chiral phase boundary, but around 1.5 along the deconfinement phase boundary.

By changing the coupling constant in the vector channel and parameter sets, the CEP of the chiral phase transition in the PNJL model can shift.
The CEPs are located at $(\mu_B^E=919.1{\rm MeV},T^E=123.5{\rm MeV})$ and $(\mu_B^E=979.5{\rm MeV},T^E=104.2{\rm MeV})$
in the PNJL-1 and PNJL-2 models, respectively. Even though the two critical baryon chemical potentials $\mu_B^c$ in the PNJL-1 and PNJL-2
are almost the same, the critical temperature in the PNJL-1 model is higher than that in the PNJL-2. For the parameters used in the PNJL-3, there
is no CEP shows up in the phase diagram. From the 3D plot in Fig. \ref{fig:Fig4-3DPNJL}, one can observe two obvious phase
boundaries for the chiral and deconfinement phase transitions in the crossover side, and the magnitude of the deconfinement ridge is much higher than
the chiral ridge. If a CEP for the chiral phase transition exists in the PNJL model, the structure of the CEP mountain for the chiral phase transition looks
as the same as that in the NJL model, and one can observe a negative region of $\k \s^2$ around the CEP above the chiral phase boundary extended
from the crossover side. The only difference is that the deconfinement phase boundary extends to the CEP mountain and merges with the
chiral phase boundary.

The structure of $\kappa \sigma^2$ along the freeze-out line in the PNJL model is more complicated due to the two separated phase boundaries. Comparing
with the NJL model, the magnitude of $\kappa \sigma^2$ at small baryon chemical potentials in the PNJL model is in agreement with experiment measurement
due to the contribution from gluodynamics.  We also choose three different freeze-out lines defined in Eq.(\ref{eq:freezeout}): 1) Starting from the back ridge of the chiral phase boundary, goes
through the negative region and then crosses the foot of the CEP mountain;  2) Starting from the back ridge of the deconfinement phase boundary,
and then crosses the foot of the CEP mountain;  3) Starting from the deconfinement phase boundary
and keeps far away from the CEP mountain. These three different freeze-out lines are indicated by long dashed, dashed-dotted and
dashed lines, respectively. Here the back/ front ridge also means the higher/lower temperature side comparing with the phase boundary, respectively.

The same as in the NJL model, the peak structure of $ \k \s^2$ along the freeze-out line in the PNJL model is solely related to the CEP mountain, when the freeze-out line crosses the CEP mountain or crosses the foot of the CEP mountain, there will be a peak showing up for $ \k \s^2$, and the location of the peak is related to the location of the CEP mountain. If there is no CEP, $ \k \s^2$ keeps flat in almost the whole chemical potential region.  In the PNJL-1 and PNJL-2 models with the existence of CEP in the phase diagram, it is found that for the first case of the freeze-out lines, one can observe the dip-peak structure for $ \k \s^2$ along the freeze-out lines. The freeze-out line starts from the back ridge of the chiral phase boundary and goes through the negative region of $\k \s^2$, then crosses the foot of the CEP mountain, therefore, $ \k \s^2$ decreases from around 0.5 and then down to negative value then rises up quickly and shows up a high peak around the critical chemical potential, thus shows a dip and then a peak structure of $ \k \s^2$ along the freeze-out line. For the second case,  the freeze-out line starts from the back ridge of the deconfinement phase boundary, and has no chance to go through the negative region, then crosses the foot of the CEP mountain.  For the third case when the freeze-out line is far away from the phase boundary, $ \k \s^2$ along the freeze-out line is almost flat.

In the PNJL-3 model, there is no CEP in the phase diagram, and no special structure of $\kappa \sigma^2$ along the freeze-out lines is observed.

\subsection{The kurtosis of the baryon number fluctuation $\kappa \sigma^2$ in the $\mu$PNJL model:}

\begin{figure}[!thb]
  \centering
  \includegraphics[width=0.9\textwidth]{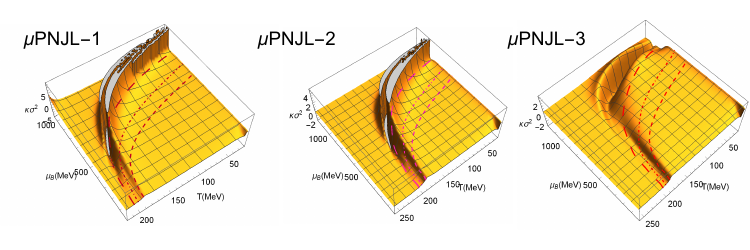}\\
  \caption{The 3D plot for the kurtosis of baryon number fluctuation $\kappa \sigma^2$ as a function of the temperature and baryon chemical potential in the
  $\mu$PNJL model. The long dashed, dashed-dotted and dashed lines in each model $\mu$PNJL-I (with I=1,2,3) are imagined freeze-out lines $\mu$pfI1, $\mu$pfI2 and $\mu$pfI3 defined in Eq.(\ref{eq:freezeout}), respectively. }
  \label{fig:Fig6-3DmuPNJL}
\end{figure}

\begin{figure}[!thb]
  \centering
  \includegraphics[width=0.9\textwidth]{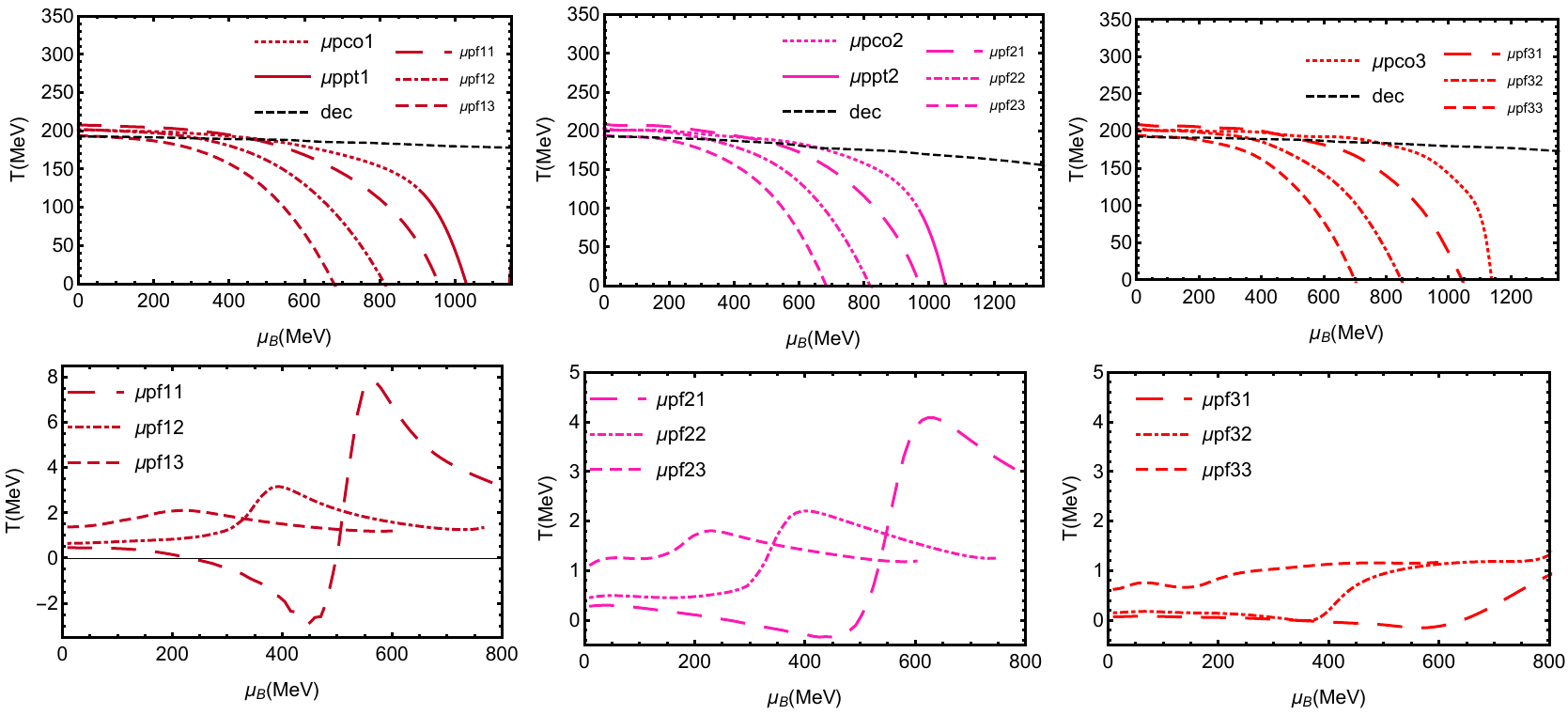}\\
  \caption{(Above) The deconfinement phase transition line indicated by short-dashed line and the chiral phase transition line in the $\mu$PNJL model with the 1st order phase transition line and the crossover indicated by solid and dotted lines, respectively. Three different freeze-out lines for each model $\mu$PNJL-I (with I=1,2,3) are defined in Eq.(\ref{eq:freezeout}) : 1) $\mu$pfI1 starts from the back ridge of the chiral phase boundary, goes through the negative region and then crosses the foot of the CEP mountain;  2) $\mu$pfI2 starts from the back ridge of the deconfinement phase boundary, and then crosses the foot of the CEP mountain;  3) $\mu$pfI3 starts from the deconfinement phase boundary
and keeps far away from the CEP mountain. These three different freeze-out lines are indicated by long dashed, dashed-dotted and
dashed lines, respectively. (Below) The 2D plot for $\kappa \sigma^2$
as a function of the baryon chemical potential in the $\mu$PNJL model along three freeze-out lines.}
  \label{fig:Fig7-2DmuPNJL}
\end{figure}

In Fig. \ref{fig:Fig6-3DmuPNJL}, we show the 3D plot for the kurtosis of baryon number fluctuation $\kappa \sigma^2$ as a function of the temperature and baryon
chemical potential in the $\mu$PNJL model. In Fig.\ref{fig:Fig7-2DmuPNJL} we show the phase transition lines and 2D plot for $\kappa \sigma^2$ as a function of the baryon chemical potential along different freeze-out lines for $\mu$PNJL-1 $\mu$PNJL-2 and $\mu$PNJL-3 models, respectively.

Same as that in the PNJL model, in the $\mu$PNJL model, there also exist two separate phase transitions for the chiral phase transition
and deconfinement phase transition, and the deconfinement phase transition line also lays below the chiral phase transition line
at small chemical potential region. From the 3D plot Fig.\ref{fig:Fig6-3DmuPNJL} one can observe two separate phase boundaries at small baryon chemical
potentials. The height of the ridge along the deconfinement phase boundary
is around 1.5 at small chemical potentials in the $\mu$PNJL model, which is similar with in the PNJL model.

Similar to that in the PNJL model, we choose three different freeze-out lines defined in Eq. (\ref{eq:freezeout}):
1) Starting from the back ridge of the chiral phase boundary, goes
through the negative region and then crosses the foot of the CEP mountain;  2) Starting from the back ridge of the deconfinement phase boundary,
and then crosses the foot of the CEP mountain;  3) Starting from the deconfinement phase boundary
and keeps far away from the CEP mountain. These three different freeze-out lines are indicated by long dashed, dashed-dotted and
dashed lines, respectively.  The structure of $\kappa \sigma^2$ along the freeze-out line in $\mu$PNJL-1 and $\mu$PNJL-2 can show the dip and peak structure
for the first case of the freeze-out lines, and $ \k \s^2$ goes to negative at the dip.

\section{The kurtosis of the baryon number fluctuation $\kappa \sigma^2$ in a realistic PNJL model}
\label{sec:realistic-model}

In last section, we have investigated gluodynamics contribution to the baryon number fluctuations, and analyzed the formation of the dip and peak structures
of the kurtosis along the imagined freeze-out lines. In this section, we will investigate the kurtosis along the experimental freeze-out line in a realistic 3-flavor PNJL model which takes into account 8-quark interaction \cite{Bhattacharyya:2016jsn}. The effective potential is given below:
\begin{eqnarray}
\Omega & =& g_S\sum_{f}{\sigma_f^2}-\frac{g_D}{2}\sigma_u\sigma_d\sigma_s
        +3\frac{g_1}{2}(\sum_{f}{\sigma_f^2})^2+3g_2\sum_f{\sigma_f^4}-6\int_0^\L \f{d^3p}{(2\pi)^3} E_f  \nonumber \\
& & -2T\int \f{d^3p}{(2\pi)^3} \ln[1+3(\Phi+\bar{\Phi}e^{-(E_f-\mu_f)/T})e^{-(E_f-\mu_f)/T}+e^{-3(E_f-\mu_f)/T}] \nonumber \\
& & -2T\int \f{d^3p}{(2\pi)^3} \ln[1+3(\bar{\Phi}+\Phi e^{-(E_f+\mu_f)/T})e^{-(E_f+\mu_f)/T}+e^{-3(E_f+\mu_f)/T}] \nonumber \\
& & +U'(\Phi,\bar{\Phi},T),
\end{eqnarray}
where $\sigma_f=\left\langle\bar{\psi}_f\psi_f \right\rangle $ corresponds to quark condensates and $f$ takes $u,d$ for two light flavors while $s$ for strange quark.  $E_f=\sqrt{p^2+M_f^2}$ with $M_f$ the dynamically generated constituent quark mass:
\be
M_f=m_f-2g_S\sigma_f+\frac{g_D}{4}\sigma_{f+1}\sigma_{f+2}-2g_1\sigma_f(\sum_{f'}{\sigma_{f'}^2})-4g_2\sigma_f^3.
\ee
If $\sigma_f=\sigma_u$, then $\sigma_{f+1}=\sigma_d$ and $\sigma_{f+2}=\sigma_s$, and so on in a clockwise manner.

$U'$ describes the contribution from self interaction of $\Phi$ and $\bar{\Phi}$ and it reads \cite{Ghosh:2007wy}:
\be
\frac{U'}{T^4}=\frac{U}{T^4}-\kappa\ln[J(\Phi,\bar{\Phi})],
\ee
where
\be
\frac{U}{T^4}=-\frac{b_2(T)}{2}\bar{\Phi}\Phi-\frac{b_3}{6}(\Phi^3+\bar{\Phi}^3)+\frac{b_4}{4}(\Phi\bar{\Phi})^2
\ee
and
\be
J=(\frac{27}{24\pi^2})(1-6\Phi\bar{\Phi}+4(\Phi^3+\bar{\Phi}^3)-3(\Phi\bar{\Phi})^2)
\ee
correspond to the effective potential of the Polyakov loop and the Jacobian of the transformation from the Polyakov loop to its trace, respectively. Besides, $\kappa$ is a dimensionless parameter. $b_2(T)$ is a temperature dependent coefficient which is chosen to have the form of
\be
b_2(T)=a_0+a_1 \frac{T_0}{T}\exp(-a_2 \frac{T}{T_0}).
\ee
Follow \cite{Bhattacharyya:2016jsn}, the parameters of the NJL part are fixed by vacuum properties and the parameters of Polyakov loop part are fixed by global fitting of the pressure density at zero chemical potential, and the details are listed in Table \ref{parameters-NJL-part} and Table \ref{parameters-Polyakov-part}, respectively.

\begin{table}[!ht]\centering
\begin{tabular}{c|c|c|c|c|c|c}\hline\centering
	$m_{u,d} ({\rm MeV})$ &  $m_s ({\rm MeV}) $ & $\Lambda ({\rm MeV})$ & $g_S\Lambda^{2}$  & $g_D\Lambda^{5}$  & $g_1 ({\rm MeV}^{-8})$     & $g_2 ({\rm MeV}^{-8})$  \\\hline
    	5.5         &   183.468    &    637.720 &    2.914          &      75.968       & $2.193\times 10^{-21}$    &  $-5.890\times 10^{-22}$\\\hline
\end{tabular}
\caption{Parameters for the NJL part in the realistic PNJL model. }
\label{parameters-NJL-part}
\end{table}

\begin{table}[!ht]\centering
	\begin{tabular}{c|c|c|c|c|c|c}\hline\centering
		$T_0$ (MeV) &  $a_0$ & $a_1$ & $a_2$  & $b_3$ & $b_4$ & $\kappa$  \\\hline
	 	  175       &  6.75  &  -9.8 & 0.26   & 0.805 & 7.555 & 0.1 \\\hline
	\end{tabular}
\caption{Parameters for the Polyakov loop part in the realistic PNJL model. }
\label{parameters-Polyakov-part}
\end{table}

\begin{figure}
 \centering
	\includegraphics[width=0.5\textwidth]{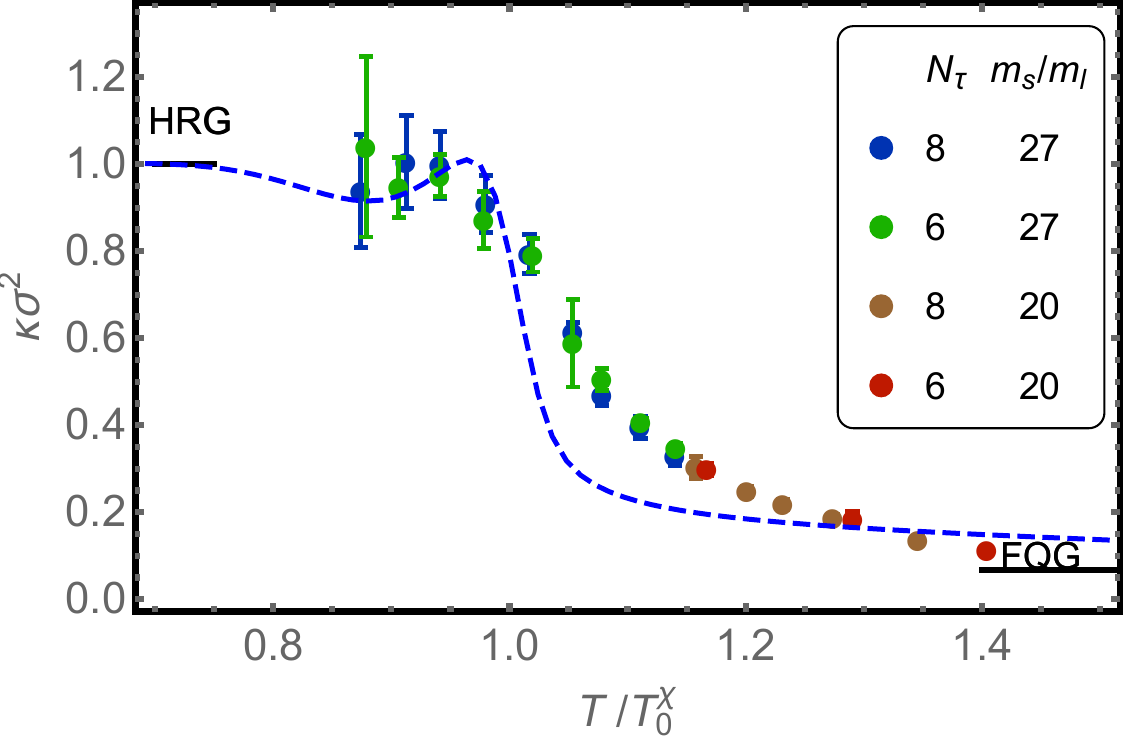}
	\caption{The kurtosis of baryon number fluctuation $\kappa\sigma^2$ in the realistic PNJL model as a function of the temperature at zero baryon number density with $T_0^{\chi}=166 {\rm MeV}$. $\kappa \sigma^2=1$ in the hadron resonance gas (HRG) limit and $\kappa \sigma^2\simeq 0.068$  in the ideal free quark gas (FQG) limit.}
\label{fig:c4}
\end{figure}

\begin{figure}[h!]
	\centering
	\includegraphics[width=0.5\textwidth]{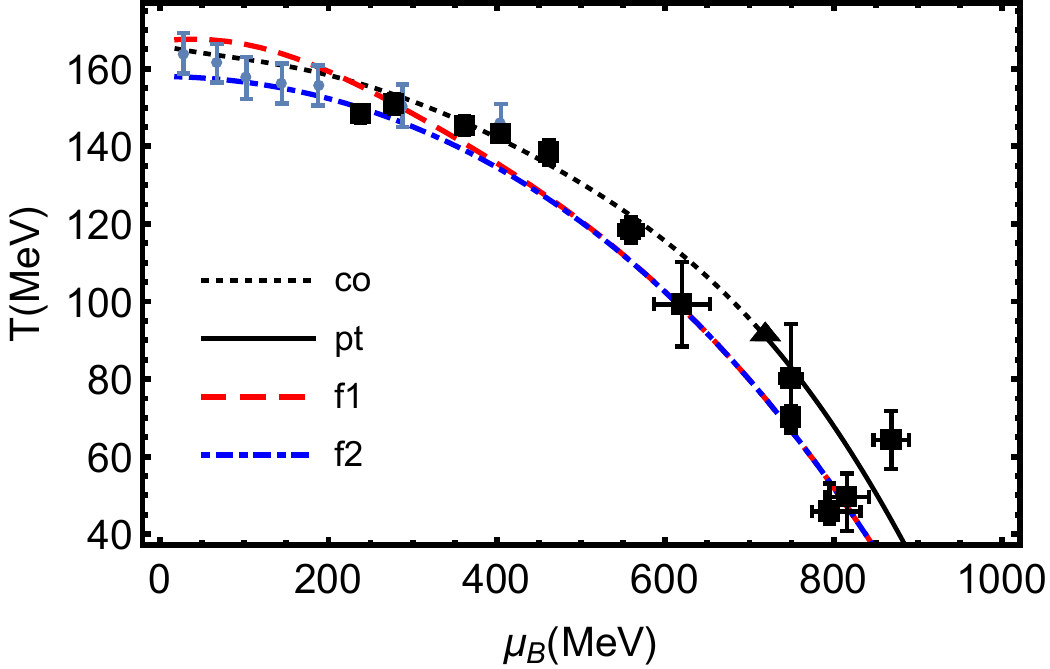}
	\caption{ The chiral phase transition line for the $u,d$ quark in the realistic PNJL model and freeze-out lines extracted from experiments. The CEP is marked by a triangle and located at $(\mu_B^E= 720 {\rm MeV}, T^E=93 {\rm MeV})$. The phase transition and crossover are shown by black line and black dotted line, respectively. The freeze-out temperatures and baryon number chemical potentials extracted from BES-I at RHIC \cite{Das:2014qca,Adamczyk:2017iwn,Kaczmarek:2017hfx} are shown in dots, and the freeze-out temperatures and baryon number chemical potentials for lower energy heavy-ion collisions summarized in \cite{Begun:2016pdy} are shown in squares, and two fitted freeze-out lines $f1, f2$ are shown by long dashed and dashed-dotted lines, respectively. }
\label{fig:phasetransition-2D-3D}
\end{figure}

With these parameters, as shown in \cite{Bhattacharyya:2016jsn}, at zero chemical potential $\mu_B=0$, the equation of state, baryon number fluctuations above the critical temperature are in good agreement with Lattice data. The kurtosis of baryon number fluctuation $\kappa\sigma^2$ as a function of the temperature at zero baryon number density is shown in Fig. \ref{fig:c4}. It is noticed that $\kappa\sigma^2$ in the realistic PNJL model  in general is in good agreement with lattice data, especially comparing with the NJL model, PNJL model as well as $\mu$PNJL model. The phase transition line is shown in Fig. \ref{fig:phasetransition-2D-3D}, and the CEP is located at $(\mu_B^E= 720 {\rm MeV}, T^E=93 {\rm MeV}) $. In Fig. \ref{fig:phasetransition-2D-3D}, the freeze-out temperatures and baryon number chemical potentials extracted from BES-I at RHIC \cite{Das:2014qca} are shown in dots, and the freeze-out temperatures and baryon number chemical potentials for lower energy heavy-ion collisions summarized in \cite{Begun:2016pdy} are shown in squares, and the two fitted freeze-out lines $f1, f2$ described by $T(\mu)=0.158-0.14\mu^2-0.04\mu^4-0.01\exp(-(\mu-0.067)/0.05)$ and $T(\mu)=0.158-0.14\mu^2-0.04\mu^4$ as used in \cite{Luo:2017faz} are shown in long dashed and  dashed-dotted lines, respectively. Note that in these two formulas the $T$ and $\mu_B$ are in $\rm{GeV}$s. The first freeze-out line $f1$ starts from the back ridge of the phase boundary and $f2$ starts from the front ridge of the phase boundary \ref{fig:phasetransition-2D-3D}.

\begin{figure}
	\centering
\includegraphics[width=0.45\textwidth]{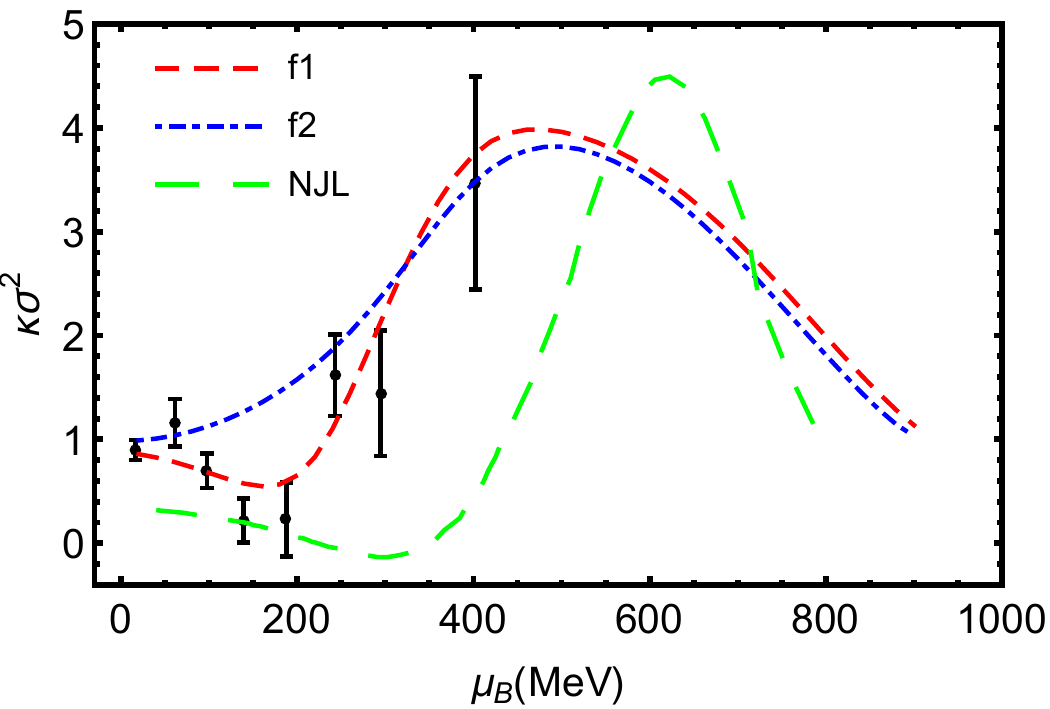}
\includegraphics[width=0.45\textwidth]{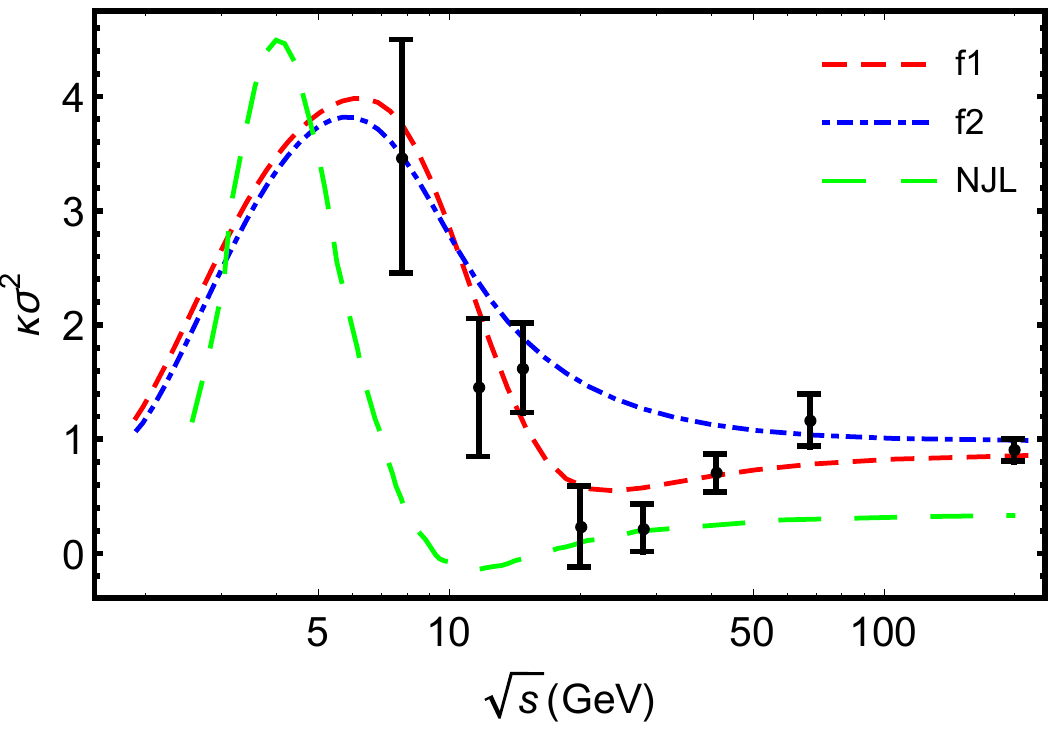}
\caption{The 2D plot for $\kappa \sigma^2$ as a function of the baryon chemical potential (Left) and the collision energy (Right) in the realistic PNJL model along freeze-out lines $f1$ (dashed line) and $f2$ (dashed-dotted line) comparing with STAR Net-proton measurement\cite{Luo:2017faz}. The long dashed line is $\kappa\sigma^2$ in a realistic NJL model \cite{Fan:2016ovc}. }
\label{fig:kurtosis-muB-energy}
\end{figure}

The kurtosis $\kappa\sigma^2$ from the realistic PNJL model along the two freeze-out lines $f1,f2$ fitted from experimental data are shown in Fig. \ref{fig:kurtosis-muB-energy}, the left figure is shown as a function of the baryon number chemical potential and the right figure is shown as a function
of the collision energy, where we have used the following relation between the chemical potential and the collision energy:
\be
\mu_{B}(\sqrt{s})=\frac{1.477}{1+0.343\sqrt{s}}.
\ee
Note that in this formula the $T$ and $\mu_B$ are also in $\rm{GeV}$s. 

We can see that the kurtosis $\kappa\sigma^2$ from the realistic PNJL model along the freeze-out line $f1$, which starts from the back ridge of the phase boundary,
develops a dip structure around $\mu_B=0.2 {\rm GeV}$($\sqrt{s}=20 {\rm GeV}$), and a peak structure at around $\mu_B=0.45 {\rm GeV}$ ($\sqrt{s}=6{\rm GeV}$). To our surprise, the kurtosis $\kappa\sigma^2$ from the realistic PNJL model along this experimental freeze-out line agree with BES-I result very well. Along the second freeze-out line $f2$, which starts from the front ridge of the phase boundary, the kurtosis only develops a peak structure at around
$\mu_B=0.45 {\rm GeV}$ ($\sqrt{s}=6{\rm GeV}$) and no dip structure is developed. This supports our qualitative analysis in Sec. \ref{sec-qualitative-model}.

\section{Conclusion and discussion}
\label{sec:Conclusion}

In this work, firstly we qualitatively investigate the kurtosis $\kappa \sigma^2$ of net baryon number fluctuation and analyze the formation of its dip and peak structures along the imagined freeze-out lines in the NJL model, PNJL model as well as $\mu$PNJL model with different parameter sets, and then we apply a realistic PNJL model and quantitatively investigate its $\kappa \sigma^2$ along the real freeze-out line extracted from experiments.

Through qualitative analysis, we find that: 1) At zero chemical potential, the magnitude of $\kappa \sigma^2$ is rather small in the NJL model comparing with lattice result, and it can reach around 1.5 in the PNJL and $\mu$PNJL models around the critical temperature. This indicates that gluodynamics plays important role in the baryon number fluctuation $\kappa \sigma^2$; 2) The peak structure of $ \k \s^2$ along the freeze-out line is solely determined by the existence of the CEP mountain. When the freeze-out line crosses the CEP mountain or crosses the foot of the CEP mountain, there will be a peak showing up for $ \k \s^2$ along the freeze-out line, and the location of the peak is close to the location of the CEP mountain. The higher the peak, the closer the peak location to the CEP. In the case of no CEP, it is found that $ \k \s^2$ keeps flat almost in the whole chemical potential region and does not show any structure; 3) The formation of the dip structure is more complicated: Firstly, it requires the existence of the CEP in the QCD phase diagram; Secondly, it is sensitive to the relation between the freeze-out line and the phase boundary. The dip structure can be formed if the freeze-out line starts from the temperature above the critical temperature at $\mu=0$ and crosses the phase boundary from above when the CEP exists in the phase diagram. Because the CEP is for chiral phase transition and there is a negative region of $\kappa \sigma^2$ from the crossover side, if the freeze-out line starts from the back ridge of the chiral phase boundary and if it goes through the negative region of $\k \s^2$, and then crosses the CEP mountain or the foot of the CEP mountain, in this case one can observe a dip structure and a peak structure, and the magnitude of $ \k \s^2$ will go to negative at the dip. If the freeze-out line starts from the back ridge of the deconfinement phase boundary, and then crosses the foot of the CEP mountain, we can also see a dip and peak structure of $ \k \s^2$ along the freeze-out line, but the magnitude of $ \k \s^2$ will not go to negative at the dip. Therefore we can read the information on how does the freeze-out line crosses the chiral phase boundary from the negative/positive value at the dip of measured $ \k \s^2$ along the freeze-out line.

Quantitatively, we use a reparameterized realistic PNJL model, with its critical temperature, equation of state and baryon number fluctuations in good agreement lattice data at zero chemical potential. To our surprise, the kurtosis $\kappa \sigma^2$ produced from the realistic PNJL model along the experimental freeze-out line agrees with BES-I data well. This may indicate that the equilibrium result can explain the BES-I data on baryon number fluctuations. Indeed, from the analysis in \cite{BraunMunzinger:2003zd}, after collision, the system reaches thermalization quickly in quite high temperature and then evolves in equilibrium state, e.g.
in the collision energy of $\sqrt{s}=200 {\rm GeV}$, the system reaches thermalization at around $T\simeq 210-230 {\rm MeV}$ , which is much higher than the freeze-out temperature as well as the phase transition temperature. It is worth to point out that the extracted freeze-out temperatures from beam energy scan measurement are indeed higher than the critical temperatures at small chemical potentials, which supports our qualitative analysis on the formation of dip structure
of $\k \s^2$ along the freeze-out line.

At last, we should mention that in this work, even though our quantitative result from static thermodynamics can describe BES-I data well,  the non-thermal effect, or memory effect \cite{Memory}, and finite size effect deserves further studies to locate the CEP.

\begin{acknowledgments}
We thank valuable discussions with H.T.Ding, W.J.Fu, X.F.Luo, J. Pawlowski, K. Redlich and G.Y.Shao This work is supported in part by the NSFC under Grant Nos. 11647173, 11725523, 11735007, 11261130311 (CRC 110 by DFG and NSFC), Chinese Academy of Sciences under Grant No. XDPB09, and the start-up funding from University of Chinese Academy of Sciences(UCAS).
\end{acknowledgments}

\end{document}